\documentclass[sigconf]{acmart}

\usepackage{booktabs} 

\usepackage{array}
\newcommand{\PreserveBackslash}[1]{\let\temp=\\#1\let\\=\temp}
\newcolumntype{C}[1]{>{\PreserveBackslash\centering}p{#1}}
\newcolumntype{R}[1]{>{\PreserveBackslash\raggedleft}p{#1}}
\newcolumntype{L}[1]{>{\PreserveBackslash\raggedright}p{#1}}

\usepackage{todonotes} 
\usepackage{amsmath}
\usepackage{bm}
\usepackage{amsfonts}
\usepackage{amsthm}
\usepackage{multirow}
\usepackage{colortbl}
\usepackage{enumitem}
\usepackage{subcaption}

\def\eqref#1{equation~\ref{#1}}

\def\1{\bm{1}}

\DeclareMathAlphabet{\mathsfit}{\encodingdefault}{\sfdefault}{m}{sl}
\SetMathAlphabet{\mathsfit}{bold}{\encodingdefault}{\sfdefault}{bx}{n}

\setcopyright{rightsretained}

\acmConference[SIGIR'19]{42nd International ACM SIGIR Conference on Research and Development in Information Retrieval}{July 2019}{Paris, France}
\acmYear{2019}
\copyrightyear{2019}

\begin{document}

\title{On the Effect of Low-Frequency Terms on Neural-IR Models}


\author{Sebastian Hofst{\"a}tter}
\affiliation{%
  \institution{TU Wien}
}
\email{s.hofstaetter@tuwien.ac.at}

\author{Navid Rekabsaz}
\affiliation{%
  \institution{Idiap Research Institute}
}
\authornote{The contribution of the second author is based upon the work supported by the Office of the Director of National Intelligence (ODNI), Intelligence Advanced Research Projects Activity (IARPA), via AFRL Contract \#FA8650-17-C-9116. The views and conclusions contained herein are those of the authors and should not be interpreted as necessarily representing the official policies or endorsements, either expressed or implied, of the ODNI, IARPA, or the U.S. Government. The U.S. Government is authorized to reproduce and distribute reprints for Governmental purposes notwithstanding any copyright annotation thereon.}
\email{navid.rekabsaz@idiap.ch}

\author{Carsten Eickhoff}
\affiliation{%
  \institution{Brown University}
}
\email{carsten@brown.edu}

\author{Allan Hanbury}
\affiliation{%
  \institution{TU Wien}
}
\email{hanbury@ifs.tuwien.ac.at}

\begin{abstract}

Low-frequency terms are a recurring challenge for information retrieval models, especially neural IR frameworks struggle with adequately capturing infrequently observed words. While these terms are often removed from neural models -- mainly as a concession to efficiency demands -- they traditionally play an important role in the performance of IR models. In this paper, we analyze the effects of low-frequency terms on the performance and robustness of neural IR models. We conduct controlled experiments on three recent neural IR models, trained on a large-scale passage retrieval collection. We evaluate the neural IR models with various vocabulary sizes for their respective word embeddings, considering different levels of constraints on the available GPU memory. 

We observe that despite the significant benefits of using larger vocabularies, the performance gap between the vocabularies can be, to a great extent, mitigated by extensive tuning of a related parameter: the number of documents to re-rank. We further investigate the use of subword-token embedding models, and in particular FastText, for neural IR models. Our experiments show that using FastText brings slight improvements to the overall performance of the neural IR models in comparison to models trained on the full vocabulary, while the improvement becomes much more pronounced for queries containing low-frequency terms. 
\end{abstract}

%
%
%
%
%

\maketitle
\vspace{-0.15cm}
\section{Introduction}
\vspace{-0.09cm}
Neural network approaches for Information Retrieval have been showing promising performance in a wide range of document retrieval tasks. Various studies apply neural methods by introducing pre-trained word embeddings into classical IR models~\cite{rekabsaz2016generalizing,rekabsaz2017word}, adapting word embeddings to retrieval tasks~\cite{Hofstaetter2019,diaz2016query}, or proposing altogether novel neural IR models~\cite{Dai2018,Xiong2017,Pang2016,Pang2017}. An essential core building block of all these approaches is the word embedding model, which defines the semantic relations between the terms. 

Typically, word embeddings are defined on a fixed vocabulary. As a common practice in neural network approaches, terms with very low collection frequencies are pruned from the vocabulary, becoming Out-Of-Vocabulary (OOV) terms. The reason for limiting the vocabulary in this way often stems from (GPU) memory constraints, efficiency considerations, or noise reduction efforts.

However, in the context of retrieval modeling, low-frequency terms are known to bear high degrees of informativeness or salience, and therefore play an important role in identifying relevant documents. In classical IR, the importance of such terms is quantified by term salience measures, such as the Inverse Document Frequency. Removing low-frequency terms in the training stage of neural IR models potentially harms the effectiveness and robustness of the derived models, especially for queries containing the affected terms. Even if neural IR models cover the full collection vocabulary, two issues remain: (1) The model performs poorly on previously unseen terms appearing at retrieval time (OOV terms). (2) Due to the lack of training data for low-frequency terms, the learned vectors may not be semantically robust.

In this study, we explore the effect of low-frequency terms on the effectiveness and robustness of neural IR models. We conduct an extensive range of controlled experiments on three recent neural IR models, namely \textit{KNRM}~\cite{Xiong2017}, \textit{CONV-KNRM}~\cite{Dai2018}, and \textit{MatchPyramid}~\cite{Pang2016}, evaluated on the MS MARCO~\cite{msmarco16} passage ranking collection, and finally propose potential solutions to the general underlying vocabulary issue in neural IR models. 

The novel contributions of this paper are two-fold: We begin by exploring the performance of neural IR models trained on different vocabularies (Section~\ref{sec:analysis}). We observe that despite the significant benefits of using larger vocabularies, model performance is highly sensitive to another essential parameter common to virtually all neural IR models: the re-ranking threshold, which defines how many of the initially retrieved documents are re-ranked by a neural IR model. We investigate the relationship between vocabulary size and re-ranking threshold, noting the sensitivity of the models to the latter, especially for models with smaller vocabularies. Our results suggest that a well-tuned re-ranking threshold can largely mitigate the negative effect of pruned vocabularies.

Secondly, we study the effect of embedding sub-word tokens in comparison to using the full vocabulary of word-level tokens (Section~\ref{sec:subword}). In particular, we investigate the use of FastText~\cite{bojanowski2017enriching}, a model based on the composition of character n-gram vector representations, designed to address OOV issues. Our results suggest that the overall performance of the model with FastText remains close to the results of using the full vocabulary. However, character-level models achieve significantly better performance on queries containing low-frequency terms. We argue that this is due to better generalization of the character-level model that benefits from other words with similar n-gram contexts. This early-stage study therefore recommends the use of sub-word token embeddings as a strategy for retaining the effectiveness and robustness of neural IR models, especially with regard to low-frequency query terms.

\vspace{-0.15cm}
\section{Background and Related Work}
\label{sec:background}
In this section, we briefly explain the sub-word embeddings, followed by discussing related work to our study. 

Sub-word embedding models produce a vector representation of a word based on composing embeddings of the character n-grams composing the word. In this way, the models can provide a semantically meaningful embedding vector even for unseen terms by exploiting the contexts of the observed terms with similar character n-grams and there are virtually no out-of-vocabulary terms. 
The FastText model~\cite{bojanowski2017enriching}, an effective and efficient sub-word embedding model, simply sums up the character n-gram vectors to build the word embedding. For highly frequent terms, FastText  directly assigns a vector per word. ELMo~\cite{Peters2018} is another well-known character-based embedding model, which in addition, takes into account the context around the word. In this work, we use FastText due its direct comparability to traditional word embeddings.

In more traditional retrieval models, Woodland et al.~\cite{woodland2000effects} explore the role of OOV terms for spoken document retrieval, proposing query and document expansion approaches. To the best of our knowledge there is no existing research on the effect of low-frequency terms on neural IR models. 

Other studies explore related aspects of neural IR models. Pyreddy et al.~\cite{Pyreddy2018} investigate the variance and consistency of kernel-based neural models over various parameter initializations. Zamani et al.~\cite{zamani2018neural} propose a method to skip the re-ranking step, and directly retrieve documents from an index of sparse representations. In contrast, in this paper, we analyze the sensitivity of the neural IR models to the re-ranking threshold parameter, since most recently proposed neural models are based on the re-ranking mechanism.

\vspace{-0.15cm}
\section{Experiment design}
\label{sec:experiment}
We conduct our experiments on the MS MARCO~\cite{msmarco16} passage re-ranking collection. The collection provides a large set of informational question-style queries from Bing's search logs, accompanied by human-annotated relevant/non-relevant passages. Besides training data, MS MARCO provides a development set -- containing queries and relevance data for evaluation -- in two sizes: sample\footnote{Provided in the form of evaluation tuples: top1000.dev.tsv} and full. In our experiments, we use the queries from the sample as our validation set and the rest of the full development set as our test set. In total, the collection consists of 8,841,822 documents, 6,980 queries for validation, and 48,598 queries for test purposes.

\vspace{-0.15cm}
\subsubsection*{Resources.} We use GloVe~\cite{pennington2014glove} word embeddings with 300 dimensions\footnote{42B lower-cased (CommonCrawl) from: \textit{\url{https://nlp.stanford.edu/projects/glove/}}}, and the FastText model, trained on the Wikipedia corpus of August 2015, with trigram-character subwords in 200 dimensions. 

We create several vocabularies based on varying thresholds to the collection frequency of terms. In our experiments, we refer to the set of terms with frequency greater or equal to $n$, as Voc-$n$. Voc-Full uses all the terms in the collection. The details of the resulting vocabularies, as well as the corresponding statistics of OOV terms, are shown in Table~\ref{tab:oov_non_basic_data}. 

\vspace{-0.1cm}
\subsubsection*{Evaluation.} We evaluate our models with the main metric for the MS MARCO ranking challenge: the Mean Reciprocal Rank measure (MRR), as well as Recall, both  at rank 10. Statistical significance tests are done using a two sided paired $t$-test ($p<0.05$).

\begin{table}[t]
    \centering
    \caption{Left: Details of the vocabularies. Right: Percentage and absolute number of test set queries with $\geq$ 1 OOV term}
    \label{tab:oov_non_basic_data}
    \vspace{-0.3cm}
    \begin{tabular}{lrrr!{\color{lightgray}\vrule}rr}
       \toprule
       \textbf{Name} & 
       \multirow{2}{*}{\textbf{\# Terms}} &\textbf{\% Covered} & \multirow{2}{*}{\textbf{Size}} & 
            \multicolumn{2}{c}{\textbf{OOV Queries}} \\
        \textbf{- Min \#} && \textbf{Terms} && \textbf{\%} & \textbf{\#} \\ \midrule
       \textbf{Voc-Full} & 3,525,473 & 100.0 &  4.23 GB & 0    & 0    \\
       \textbf{Voc-5}    & 542,878   & 15.4  &  651 MB & 1.23  & 596  \\
       \textbf{Voc-10}   & 314,607   & 8.9   &  378 MB & 1.98  & 962  \\
       \textbf{Voc-25}   & 169,983   & 4.8   &  204 MB & 5.07  & 2464  \\
       \textbf{Voc-50}   & 111,815   & 3.2   &  134 MB & 8.06  & 3917  \\ 
       \textbf{Voc-100}  & 75,805    & 2.2   &  91  MB & 12.07 & 5864  \\ 
       \midrule
       \textbf{FastText} & 2,950,302 & 100.0  &  2.36 GB & 0 & 0  \\ 
        \bottomrule
    \vspace{-1.0cm}
    \end{tabular}
\end{table}

\vspace{-0.1cm}
\subsubsection*{Neural Retrieval Models.} \textbf{\emph{KNRM}}~\cite{Xiong2017} establishes a similarity matching matrix using the embeddings of query and document terms. The model then estimates the relevance score based on the outputs of a set of Gaussian kernel functions, applied on the matching matrix. 

\noindent
\textbf{\emph{CONV-KNRM}}~\cite{Dai2018} extends \textit{KNRM} by adding a Convolutional Neural Network (CNN) layer on top of the word embedding matrix, enabling learning word-level n-gram representations.

\noindent
\textbf{\emph{MatchPyramid}}~\cite{Pang2016} ranking model is inspired by deep neural image processing architectures. Similar to \textit{KNRM}, the model first computes the similarity matching matrix, which is used for several stacked layers of CNN with dynamic max-pooling. Like the two previous models, this architecture facilitates end-to-end training.

\vspace{-0.1cm}
\subsubsection*{Implementation and Parameter Setting}
We use the Anserini~\cite{Yang2017} toolkit to compute the \textit{BM25}, and \textit{RM3} models. The model parameters are tuned on the validation set, resulting in $k_1 = 0.6$, $b = 0.8$ for \textit{BM25}. We observe no significant performance increase for \textit{RM3}, therefore we only report \textit{BM25} baseline results. The \textit{BM25} rankings are used as a starting point for the neural re-ranking models. 

We implement the neural models in PyTorch~\cite{pytorch2017}.\footnote{Our code is available at \textit{\url{https://github.com/sebastian-hofstaetter/sigir19-neural-ir}}} We project all characters to lower case and apply tokenization using the WordTokenizer provided by AllenNLP~\cite{Gardner2017AllenNLP}. We use the Adam optimizer with learning rate 0.001, 1 epoch, and early stopping. We use a batch size of 64, and the maximum word length of queries and documents is set to 30 and 180, respectively. In all models, the pre-trained word embeddings are updated during training. 

Regarding model-specific parameters, for \textit{KNRM} and \textit{CONV-KNRM}, we set the number of kernels to $11$ with the mean values of the Gaussian kernels varying from $-1$ to $+1$ in steps of $0.2$ (one extra kernel is added for exact matching), and standard deviation of $0.1$ for all kernels. The dimension of the CNN vectors in \textit{CONV-KNRM} is set to $128$. In the MatchPyramid model, we set the number of convolution layers to $5$, each with kernel size $3\times3$ and 16 convolution channels. For each model, we find the best re-ranking threshold parameter by extensively tuning it on a range from $1$ to $300$, based on the MRR results of the validation set.

\begin{table*}[t!]
    \centering
    \caption{Evaluation results using different vocabulary sizes \& best re-ranking threshold on the validation set.\\Top: best performance on the validation set. Bottom: results on the test set. The best results per model are shown in bold}
    \label{tab:all_results}
    \vspace{-0.3cm}
    \setlength\tabcolsep{4.6pt}
    \begin{tabular}{cl!{\color{lightgray}\vrule}rr!{\color{lightgray}\vrule}rr!{\color{lightgray}\vrule}rr!{\color{lightgray}\vrule}rr!{\color{lightgray}\vrule}rr!{\color{lightgray}\vrule}rr!{\color{lightgray}\vrule}rr}
       \toprule
       &\multirow{2}{*}{\textbf{Model}}&
       \multicolumn{2}{c!{\color{lightgray}\vrule}}{\textbf{Voc-100}}&
       \multicolumn{2}{c!{\color{lightgray}\vrule}}{\textbf{Voc-50}}&
       \multicolumn{2}{c!{\color{lightgray}\vrule}}{\textbf{Voc-25}}&
       \multicolumn{2}{c!{\color{lightgray}\vrule}}{\textbf{Voc-10}}&
       \multicolumn{2}{c!{\color{lightgray}\vrule}}{\textbf{Voc-5}}&
       \multicolumn{2}{c!{\color{lightgray}\vrule}}{\textbf{Voc-Full}}&
       \multicolumn{2}{c}{\textbf{FastText}}\\
       && MRR & Recall & MRR & Recall & MRR & Recall & MRR & Recall & MRR & Recall & MRR & Recall & MRR & Recall \\
        \midrule
        \parbox[t]{2mm}{\multirow{3}{*}{\rotatebox[origin=c]{90}{\small\textbf{Val.}}}} &
        \textit{\textbf{MatchPyramid}}      & 0.220 & 0.423 & 0.218 & 0.427 & 0.224 & 0.442 & 0.234 & 0.446 & 0.231 & 0.459 & 0.239 & 0.467 & \textbf{0.245} & 0.477 \\ 
        &\textit{\textbf{KNRM}}             & 0.209 & 0.404 & 0.209 & 0.404 & 0.209 & 0.404 & 0.221 & 0.454 & 0.224 & 0.457 & \textbf{0.232} & \textbf{0.467} & 0.230 & 0.456 \\
        &\textit{\textbf{CONV-KNRM}}        & 0.253 & 0.469 & 0.249 & 0.474 & 0.256 & 0.480 & 0.261 & 0.492 & 0.266 & 0.504 & 0.276 & 0.526 & \textbf{0.278} & 0.519 \\  

        \midrule
        \parbox[t]{2mm}{\multirow{3}{*}{\rotatebox[origin=c]{90}{\small\textbf{Test}}}} &
        \textit{\textbf{MatchPyramid}}      & 0.223 & 0.426 & 0.227 & 0.431 & 0.221 & 0.442 & 0.237 & 0.458 & 0.232 & 0.458 & 0.239 & 0.465 &  \textbf{0.247} & \textbf{0.472} \\ 
        &\textit{\textbf{KNRM}}             & 0.210 & 0.407 & 0.211 & 0.407 & 0.211 & 0.407 & 0.222 & 0.453 & 0.225 & 0.455 & \textbf{0.235} & \textbf{0.468} & 0.227 & 0.451 \\
        &\textit{\textbf{CONV-KNRM}}        & 0.248 & 0.472 & 0.249 & 0.470 & 0.250 & 0.481 & 0.260 & 0.488 & 0.264 & 0.503 & 0.273 & 0.518 & \textbf{0.275} & \textbf{0.519} \\ 

        \bottomrule
    \end{tabular}
\end{table*}

\begin{figure*}[t]
    \centering
    \subcaptionbox{\textit{MatchPyramid}}{\includegraphics[trim={0.5cm 0.5cm 0.5cm 0.5cm} ,width=0.32\textwidth]{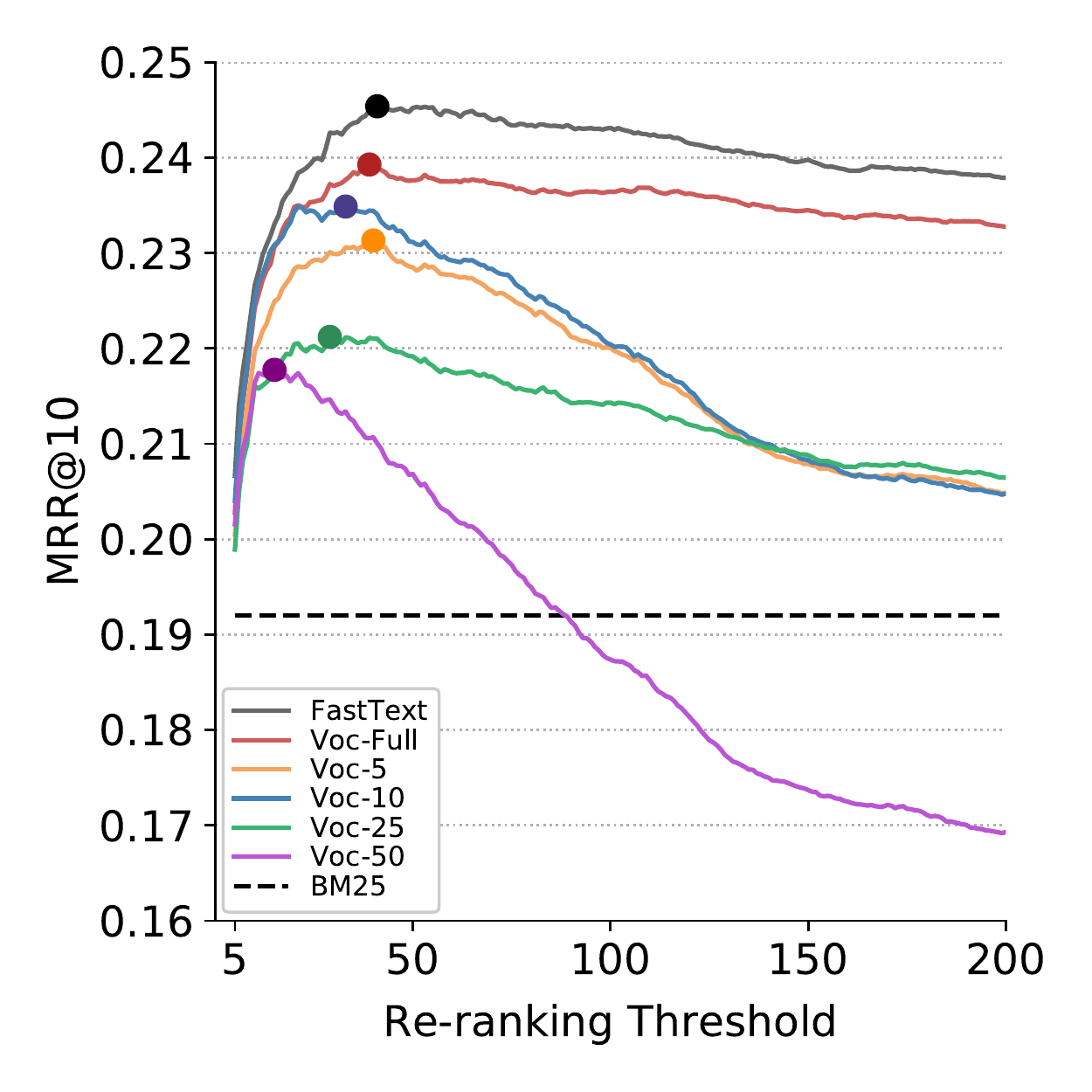}}%
    \hfill
    \subcaptionbox{\textit{KNRM}}{\includegraphics[trim={0.5cm 0.5cm 0.5cm 0.5cm} ,width=0.32\textwidth]{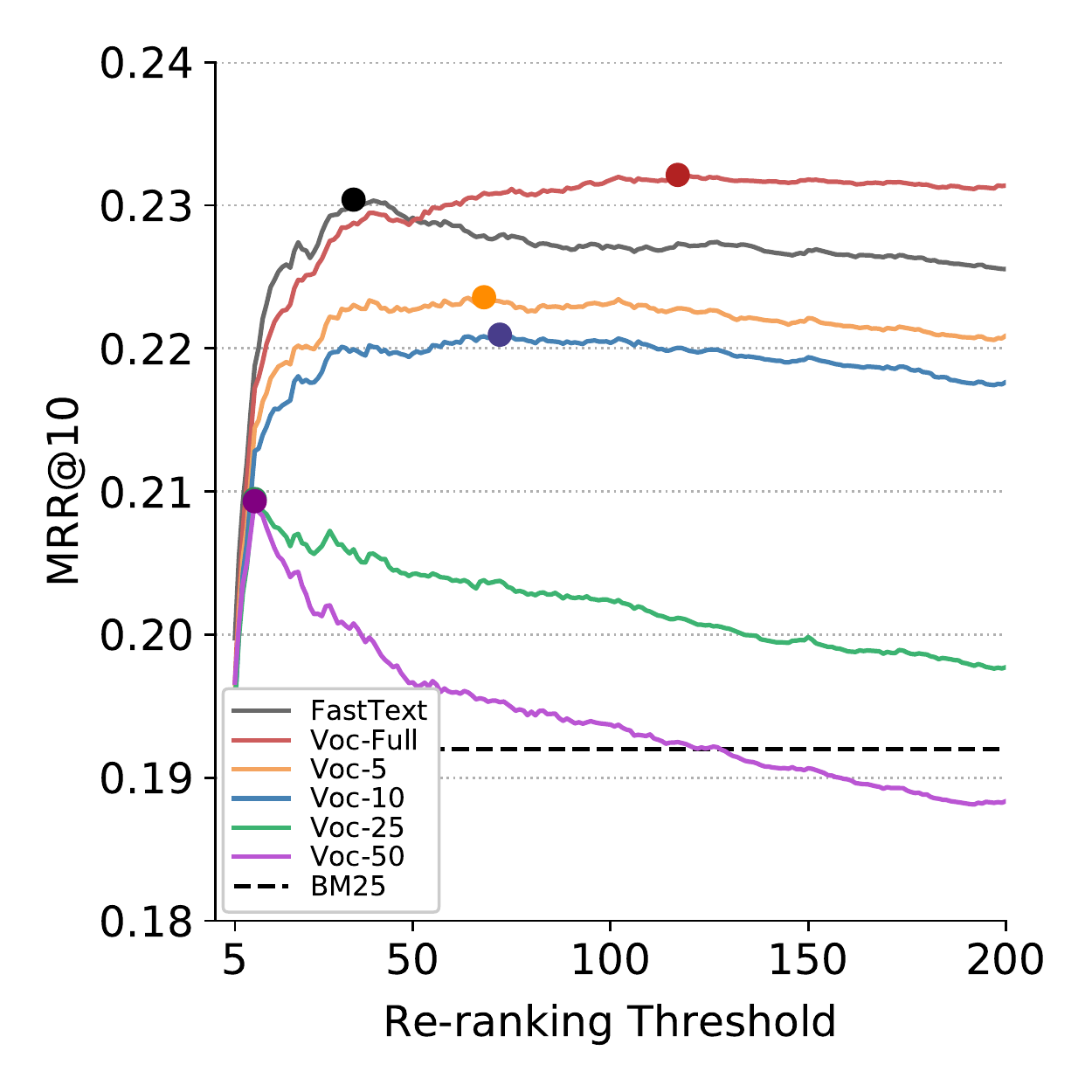}}%
    \hfill
    \subcaptionbox{\textit{CONV-KNRM}}{\includegraphics[trim={0.5cm 0.5cm 0.5cm 0.5cm} ,width=0.32\textwidth]{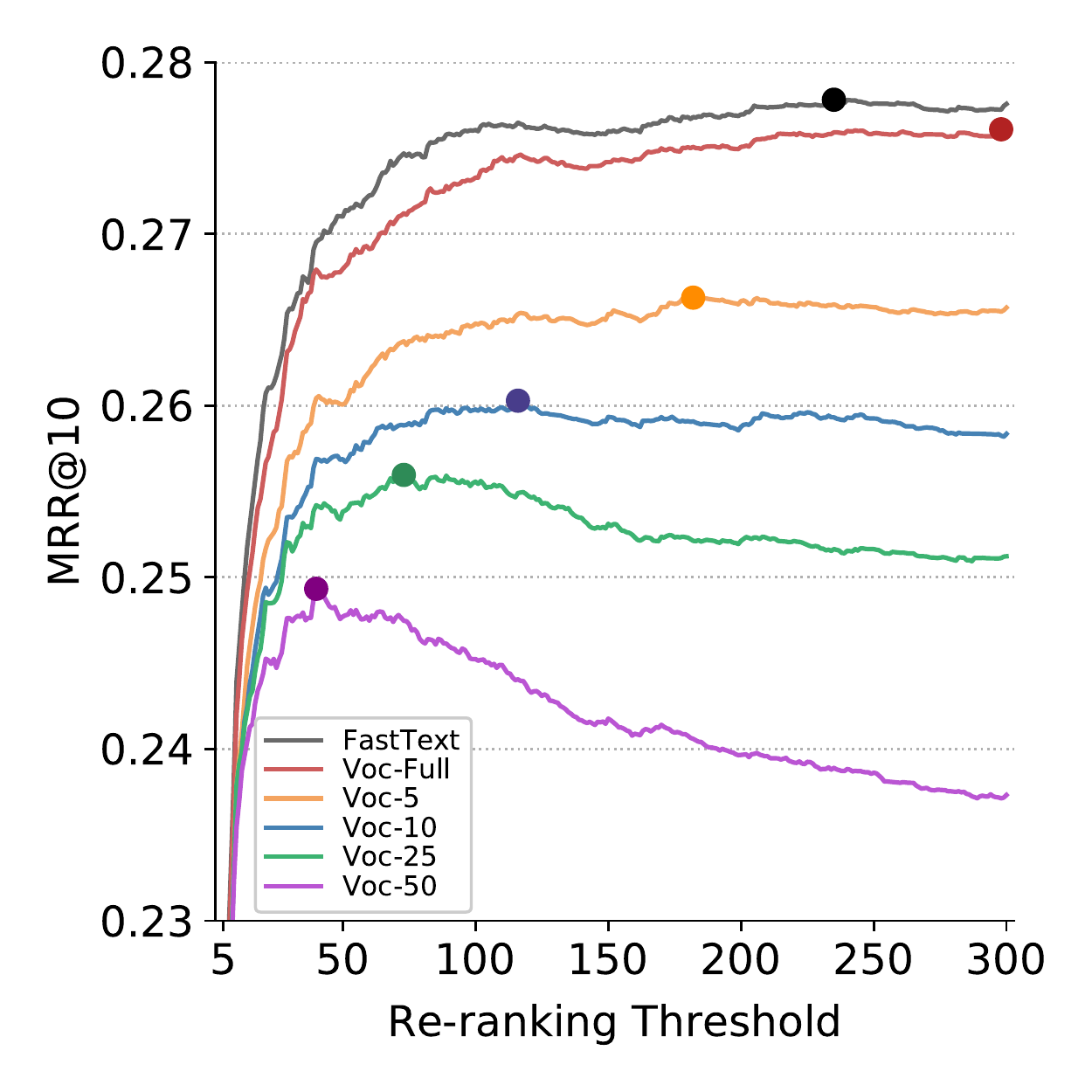}}%
    \centering
    \vspace{-0.2cm}
    \centering
    \caption{Sensitivity of the models to the re-ranking threshold parameter for different vocabularies. The best performing parameter setting are indicated on the plots.}
    \label{fig:cs_at_n_plots}
    \vspace{-0.2cm}
\end{figure*}
\section{Effect of the Vocabulary Size}
\label{sec:analysis}
The performance of the neural ranking models, trained on various vocabularies as well as on FastText embeddings, on both validation and test sets are shown in Table \ref{tab:all_results}. We calculate tests of significance between the pairs of rankings, and mention the results in the following. We also evaluate \textit{BM25}, achieving an MRR of $0.192$ and Recall of $0.407$ on the test set. Consistent with previous studies, the \textit{BM25} model is outperformed by all neural ranking models, and \textit{CONV-KNRM} shows the best overall performance~\cite{Dai2018,Xiong2017,Pang2016}.

Comparing the results over each model, in two out of the three models, the FastText embedding significantly outperforms Voc-Full, while FastText only requires 55\% of the memory needed by Voc-Full (based on the statistics in Table~\ref{tab:oov_non_basic_data}). Looking at the results of the models with various vocabulary sizes, using Voc-Full brings significant advantages in comparison to using smaller vocabularies. However, their differences become marginal, especially for the models with Voc-5 and Voc-10 vocabulary sets, taking into account that the embeddings of the Voc-5 and Voc-10 vocabularies require much less memory space, namely only 15\% (Voc-5) and 8\% (Voc-10) of the memory used by the Voc-Full embeddings.

While the reported results are based on an exhaustive tuning of hyper-parameters on the validation set, in the following we study the sensitivity of the models to the re-ranking threshold, an important -- but not well studied -- hyper-parameter of the neural IR models. Figure~\ref{fig:cs_at_n_plots} demonstrates the sensitivity of the three neural IR models to the changes of the re-ranking threshold parameter. Looking at the trends in the plots, as the performance improves, either by using a better performing model or a bigger vocabulary size, the models become less sensitive to the re-ranking threshold. Such that the optimal re-ranking thresholds also become larger, indicating that the model is able to effectively generalize over a larger set of non-relevant documents. Since increasing the re-ranking threshold mostly adds non-relevant documents. On the other hand, models with lower performances (\textit{MatchPyramid} and \textit{KNRM}), especially with smaller vocabularies, are highly sensitive to the re-ranking threshold. For such models, an exhaustive parameter search provides a significant enhancement. This indicates the importance of well-tuning the re-ranking threshold parameter, especially in scenarios with constrained memory resources.

Finally, to confirm whether the effects of tuning the re-ranking threshold on the validation set is also transferred to the test set, we compare the results on the validation and test set in Table~\ref{tab:all_results}. As shown, even on models with high sensitivity to the re-ranking threshold, the results are highly similar, indicating the effectiveness of extensive tuning of the re-ranking threshold\footnote{We should note that the cause of this effect might be due to high underlying similarities between the validation and test set of the particular collection. However, further investigations on this aspect is out of the scope of this paper.}. 

\begin{figure}[t]
   \includegraphics[trim={0.17cm 0.3cm 0cm 0.3cm} ,
   width=0.478\textwidth]{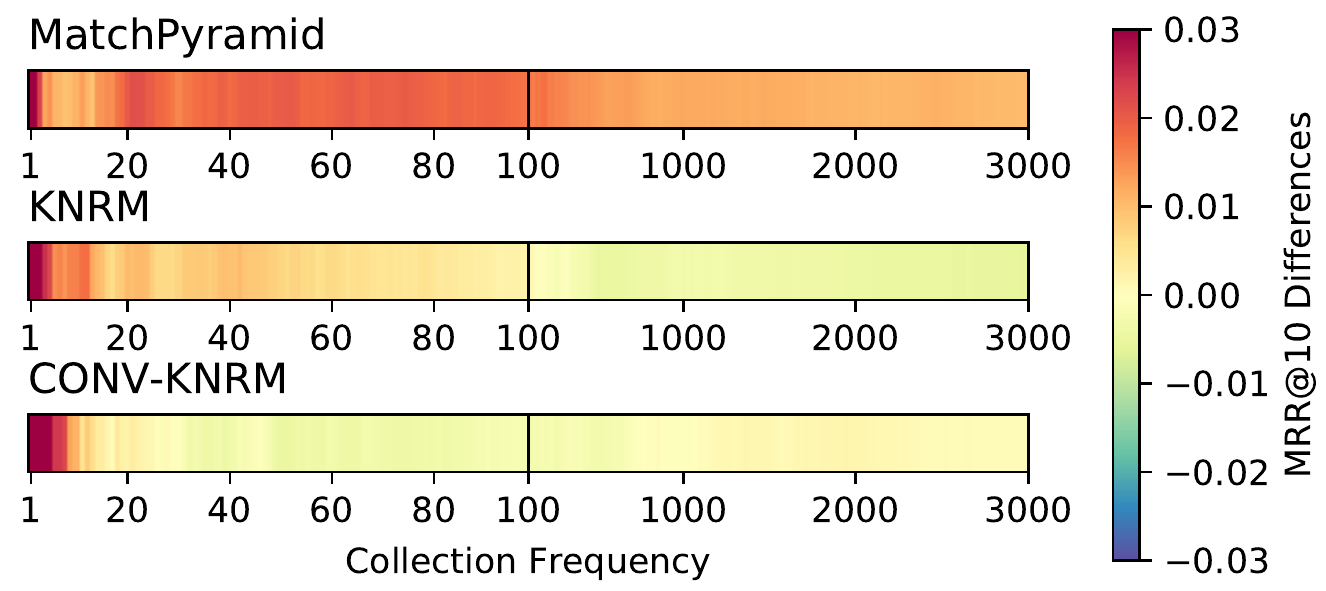}
    \centering
    \caption{MRR differences of the models, trained on the FastText embeddings and the embeddings with full vocabularies, over the queries with minimum collection frequency of their terms smaller or equal to the X-axis\\ \textit{(red = FastText is better, blue=Vocab-Full is better)}}
    \label{fig:diff_heatmap}
    \vspace{-0.4cm}
\end{figure}
\section{Queries with Low-Frequency Terms}
\label{sec:subword}

In this section, we take a closer look at the differences between the models trained on the traditional embeddings (GloVe in our experiments) using different vocabularies, and the ones trained on the FastText embeddings. 

Figure \ref{fig:diff_heatmap} shows the MRR differences of the neural ranking models, using the traditional embeddings with the Voc-Full vocabularies, to the ones using FastText, over the range of collection frequencies. For each point on the X axes, we calculate the MRR values for the queries, which at least have one term with collection frequency of equal to or smaller than the corresponding value of that point. The figure reveals strong contrast between the area, related to the queries with very low-frequency terms, and the rest, indicating higher performances of the models with FastText for these queries. 

\begin{figure}[t]
   \includegraphics[trim={0.2cm 0.5cm 0.2cm 0.4cm} ,
   width=0.475\textwidth]{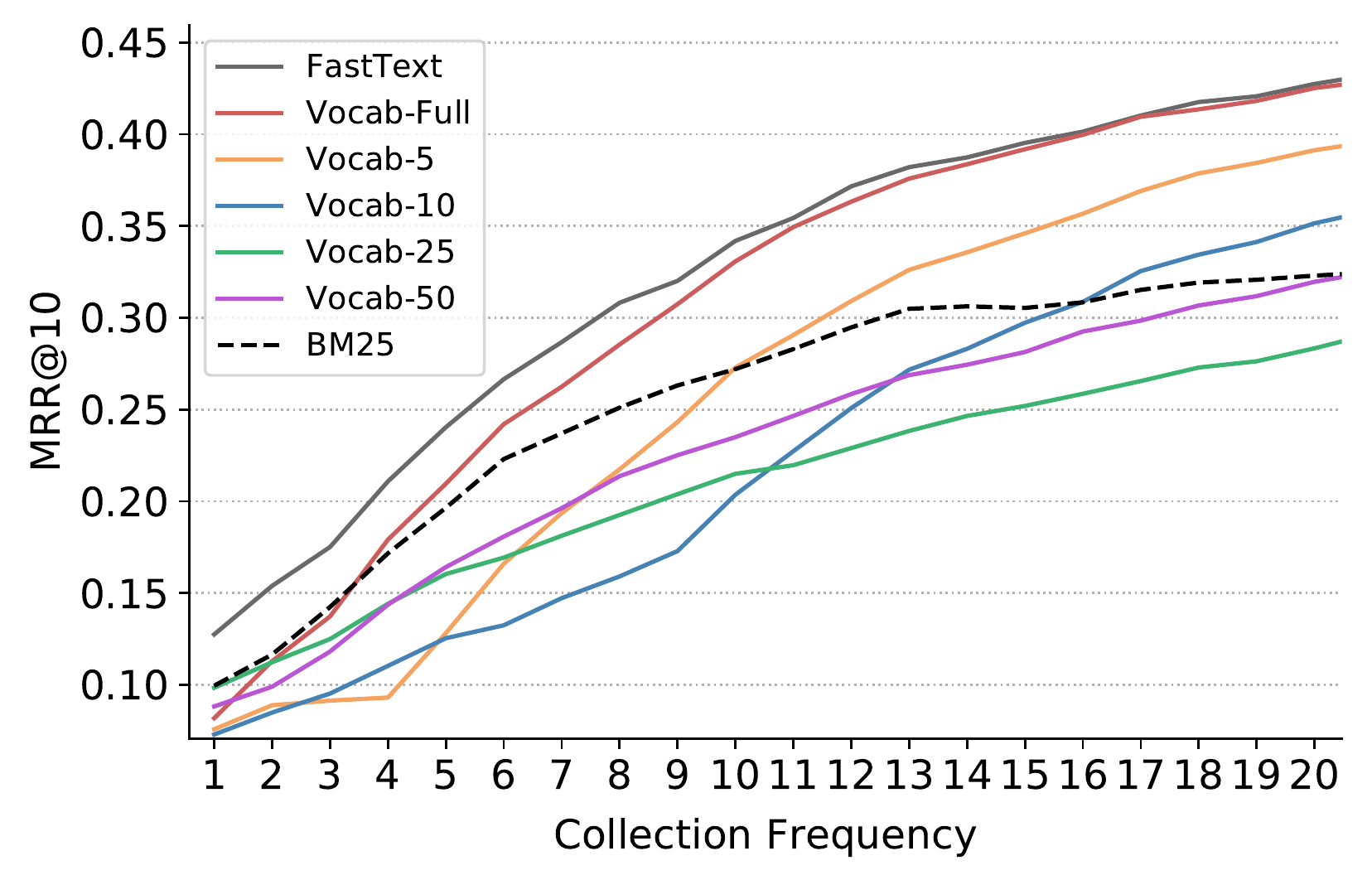}
    \centering
    \caption{MRR results of \textit{CONV-KNRM} over the queries with at least one term with collection frequency smaller than or equal to the values on the X-axis}
    \label{fig:diff_plot}
    \vspace{-0.5cm}
\end{figure}

Let us have a closer look at this area. Figure~\ref{fig:diff_plot} shows the MRR of the \textit{CONV-KNRM} models, using the traditional word embeddings with different vocabularies, as well as the one using the FastText embeddings, for queries with very infrequent terms. The MRR values are calculated in the same fashion as in Figure~\ref{fig:diff_heatmap}.

As shown, the model with FastText by a large margin improves all other models, especially until a collection frequency of around 10 to 15. Interestingly, BM25, as an exact term matching model, shows better performance than neural IR models with traditional embeddings, especially on very low values. We argue that the low performance of the models with traditional embeddings is due to the lack of enough contexts for learning meaningful representations, which causes ineffective semantic similarity estimations. On the other hand, the subword embeddings exploit the contexts of other observed terms in the collection with similar character n-grams. Therefore, the neural ranking models with subword embeddings still benefit from meaningful semantic relations between very infrequent words, outperforming the ranking models based on traditional embeddings as well as exact matching.

\section{Conclusion}
Our work takes a first step to understanding the effects of infrequent terms in neural ranking models, and exploit novel representation learning approaches to address it. We first study the sensitivity of the neural IR models to their vocabulary size, pointing out the importance of fine-grained tuning of the re-ranking threshold. We then investigate the effects of using subword embeddings in neural IR models, showing that using these embeddings in particular brings remarkable improvements to the performance of queries containing very low-frequency terms. As future work, we aim to pursue the investigations of this study into the area of query performance prediction of neural IR models.
\vspace{-0.1cm}

\bibliographystyle{ACM-Reference-Format}
\bibliography{my-references}


\begin{thebibliography}{00}


\ifx \showCODEN    \undefined \def \showCODEN     #1{\unskip}     \fi
\ifx \showDOI      \undefined \def \showDOI       #1{#1}\fi
\ifx \showISBNx    \undefined \def \showISBNx     #1{\unskip}     \fi
\ifx \showISBNxiii \undefined \def \showISBNxiii  #1{\unskip}     \fi
\ifx \showISSN     \undefined \def \showISSN      #1{\unskip}     \fi
\ifx \showLCCN     \undefined \def \showLCCN      #1{\unskip}     \fi
\ifx \shownote     \undefined \def \shownote      #1{#1}          \fi
\ifx \showarticletitle \undefined \def \showarticletitle #1{#1}   \fi
\ifx \showURL      \undefined \def \showURL       {\relax}        \fi
\providecommand\bibfield[2]{#2}
\providecommand\bibinfo[2]{#2}
\providecommand\natexlab[1]{#1}
\providecommand\showeprint[2][]{arXiv:#2}

\bibitem[\protect\citeauthoryear{Bajaj, Campos, Craswell, Deng, Gao, Liu,
  Majumder, Mcnamara, Mitra, and Nguyen}{Bajaj et~al\mbox{.}}{2016}]%
        {msmarco16}
\bibfield{author}{\bibinfo{person}{Payal Bajaj}, \bibinfo{person}{Daniel
  Campos}, \bibinfo{person}{Nick Craswell}, \bibinfo{person}{Li Deng},
  \bibinfo{person}{Jianfeng Gao}, \bibinfo{person}{Xiaodong Liu},
  \bibinfo{person}{Rangan Majumder}, \bibinfo{person}{Andrew Mcnamara},
  \bibinfo{person}{Bhaskar Mitra}, {and} \bibinfo{person}{Tri Nguyen}.}
  \bibinfo{year}{2016}\natexlab{}.
\newblock \showarticletitle{{MS MARCO : A Human Generated MAchine Reading
  COmprehension Dataset}}. In \bibinfo{booktitle}{{\em Proc. of NIPS}}.
\newblock


\bibitem[\protect\citeauthoryear{Bojanowski, Grave, Joulin, and
  Mikolov}{Bojanowski et~al\mbox{.}}{2017}]%
        {bojanowski2017enriching}
\bibfield{author}{\bibinfo{person}{Piotr Bojanowski}, \bibinfo{person}{Edouard
  Grave}, \bibinfo{person}{Armand Joulin}, {and} \bibinfo{person}{Tomas
  Mikolov}.} \bibinfo{year}{2017}\natexlab{}.
\newblock \showarticletitle{Enriching Word Vectors with Subword Information}.
\newblock \bibinfo{journal}{{\em Tr. of the ACL\/}}  \bibinfo{volume}{5}
  (\bibinfo{year}{2017}).
\newblock


\bibitem[\protect\citeauthoryear{Dai, Xiong, Callan, and Liu}{Dai
  et~al\mbox{.}}{2018}]%
        {Dai2018}
\bibfield{author}{\bibinfo{person}{Zhuyun Dai}, \bibinfo{person}{Chenyan
  Xiong}, \bibinfo{person}{Jamie Callan}, {and} \bibinfo{person}{Zhiyuan Liu}.}
  \bibinfo{year}{2018}\natexlab{}.
\newblock \showarticletitle{{Convolutional Neural Networks for Soft-Matching
  N-Grams in Ad-hoc Search}}. In \bibinfo{booktitle}{{\em Proc. of WSDM}}.
\newblock


\bibitem[\protect\citeauthoryear{Diaz, Mitra, and Craswell}{Diaz
  et~al\mbox{.}}{2016}]%
        {diaz2016query}
\bibfield{author}{\bibinfo{person}{Fernando Diaz}, \bibinfo{person}{Bhaskar
  Mitra}, {and} \bibinfo{person}{Nick Craswell}.}
  \bibinfo{year}{2016}\natexlab{}.
\newblock \showarticletitle{Query Expansion with Locally-Trained Word
  Embeddings}. In \bibinfo{booktitle}{{\em In Proc. of ACL}}.
\newblock


\bibitem[\protect\citeauthoryear{Gardner, Grus, Neumann, Tafjord, Dasigi, Liu,
  Peters, Schmitz, and Zettlemoyer}{Gardner et~al\mbox{.}}{2017}]%
        {Gardner2017AllenNLP}
\bibfield{author}{\bibinfo{person}{Matt Gardner}, \bibinfo{person}{Joel Grus},
  \bibinfo{person}{Mark Neumann}, \bibinfo{person}{Oyvind Tafjord},
  \bibinfo{person}{Pradeep Dasigi}, \bibinfo{person}{Nelson~F. Liu},
  \bibinfo{person}{Matthew Peters}, \bibinfo{person}{Michael Schmitz}, {and}
  \bibinfo{person}{Luke~S. Zettlemoyer}.} \bibinfo{year}{2017}\natexlab{}.
\newblock \showarticletitle{AllenNLP: A Deep Semantic Natural Language
  Processing Platform}.
\newblock


\bibitem[\protect\citeauthoryear{Hofst{\"a}tter, Rekabsaz, Lupu, Eickhoff, and
  Hanbury}{Hofst{\"a}tter et~al\mbox{.}}{2019}]%
        {Hofstaetter2019}
\bibfield{author}{\bibinfo{person}{Sebastian Hofst{\"a}tter},
  \bibinfo{person}{Navid Rekabsaz}, \bibinfo{person}{Mihai Lupu},
  \bibinfo{person}{Carsten Eickhoff}, {and} \bibinfo{person}{Allan Hanbury}.}
  \bibinfo{year}{2019}\natexlab{}.
\newblock \showarticletitle{Enriching Word Embeddings for Patent Retrieval with
  Global Context}. In \bibinfo{booktitle}{{\em Proc. of ECIR}}.
\newblock


\bibitem[\protect\citeauthoryear{Pang, Lan, Guo, Xu, and Cheng}{Pang
  et~al\mbox{.}}{2017}]%
        {Pang2017}
\bibfield{author}{\bibinfo{person}{Liang Pang}, \bibinfo{person}{Yanyan Lan},
  \bibinfo{person}{Jiafeng Guo}, \bibinfo{person}{Jun Xu}, {and}
  \bibinfo{person}{Xueqi Cheng}.} \bibinfo{year}{2017}\natexlab{}.
\newblock \showarticletitle{{A Deep Investigation of Deep IR Models}}. In
  \bibinfo{booktitle}{{\em Proc. of SIGIR Neu-IR'17}}.
\newblock


\bibitem[\protect\citeauthoryear{Pang, Lan, Guo, Xu, Wan, and Cheng}{Pang
  et~al\mbox{.}}{2016}]%
        {Pang2016}
\bibfield{author}{\bibinfo{person}{Liang Pang}, \bibinfo{person}{Yanyan Lan},
  \bibinfo{person}{Jiafeng Guo}, \bibinfo{person}{Jun Xu},
  \bibinfo{person}{Shengxian Wan}, {and} \bibinfo{person}{Xueqi Cheng}.}
  \bibinfo{year}{2016}\natexlab{}.
\newblock \showarticletitle{{Text Matching as Image Recognition}}. In
  \bibinfo{booktitle}{{\em Proc of. AAAI}}.
\newblock


\bibitem[\protect\citeauthoryear{Paszke, Gross, Chintala, Chanan, Yang, DeVito,
  Lin, Desmaison, Antiga, and Lerer}{Paszke et~al\mbox{.}}{2017}]%
        {pytorch2017}
\bibfield{author}{\bibinfo{person}{Adam Paszke}, \bibinfo{person}{Sam Gross},
  \bibinfo{person}{Soumith Chintala}, \bibinfo{person}{Gregory Chanan},
  \bibinfo{person}{Edward Yang}, \bibinfo{person}{Zachary DeVito},
  \bibinfo{person}{Zeming Lin}, \bibinfo{person}{Alban Desmaison},
  \bibinfo{person}{Luca Antiga}, {and} \bibinfo{person}{Adam Lerer}.}
  \bibinfo{year}{2017}\natexlab{}.
\newblock \showarticletitle{Automatic differentiation in PyTorch}. In
  \bibinfo{booktitle}{{\em NIPS-W}}.
\newblock


\bibitem[\protect\citeauthoryear{Pennington, Socher, and Manning}{Pennington
  et~al\mbox{.}}{2014}]%
        {pennington2014glove}
\bibfield{author}{\bibinfo{person}{Jeffrey Pennington},
  \bibinfo{person}{Richard Socher}, {and} \bibinfo{person}{Christopher
  Manning}.} \bibinfo{year}{2014}\natexlab{}.
\newblock \showarticletitle{Glove: Global vectors for word representation}. In
  \bibinfo{booktitle}{{\em In Proc of EMNLP}}.
\newblock


\bibitem[\protect\citeauthoryear{Peters, Neumann, Iyyer, Gardner, Clark, Lee,
  and Zettlemoyer}{Peters et~al\mbox{.}}{2018}]%
        {Peters2018}
\bibfield{author}{\bibinfo{person}{Matthew~E Peters}, \bibinfo{person}{Mark
  Neumann}, \bibinfo{person}{Mohit Iyyer}, \bibinfo{person}{Matt Gardner},
  \bibinfo{person}{Christopher Clark}, \bibinfo{person}{Kenton Lee}, {and}
  \bibinfo{person}{Luke Zettlemoyer}.} \bibinfo{year}{2018}\natexlab{}.
\newblock \showarticletitle{{Deep contextualized word representations}}. In
  \bibinfo{booktitle}{{\em Proc. of NAACL}}.
\newblock


\bibitem[\protect\citeauthoryear{Pyreddy, Ramaseshan, Joshi, Dai, Xiong,
  Callan, and Liu}{Pyreddy et~al\mbox{.}}{2018}]%
        {Pyreddy2018}
\bibfield{author}{\bibinfo{person}{Mary~Arpita Pyreddy},
  \bibinfo{person}{Varshini Ramaseshan}, \bibinfo{person}{Narendra~Nath Joshi},
  \bibinfo{person}{Zhuyun Dai}, \bibinfo{person}{Chenyan Xiong},
  \bibinfo{person}{Jamie Callan}, {and} \bibinfo{person}{Zhiyuan Liu}.}
  \bibinfo{year}{2018}\natexlab{}.
\newblock \showarticletitle{{Consistency and Variation in Kernel Neural Ranking
  Model}}. In \bibinfo{booktitle}{{\em Proc. of SIGIR}}.
\newblock


\bibitem[\protect\citeauthoryear{Rekabsaz, Lupu, Hanbury, and Zamani}{Rekabsaz
  et~al\mbox{.}}{2017}]%
        {rekabsaz2017word}
\bibfield{author}{\bibinfo{person}{Navid Rekabsaz}, \bibinfo{person}{Mihai
  Lupu}, \bibinfo{person}{Allan Hanbury}, {and} \bibinfo{person}{Hamed
  Zamani}.} \bibinfo{year}{2017}\natexlab{}.
\newblock \showarticletitle{Word Embedding Causes Topic Shifting; Exploit
  Global Context!}. In \bibinfo{booktitle}{{\em In Proc. of SIGIR}}.
\newblock


\bibitem[\protect\citeauthoryear{Rekabsaz, Lupu, Hanbury, and Zuccon}{Rekabsaz
  et~al\mbox{.}}{2016}]%
        {rekabsaz2016generalizing}
\bibfield{author}{\bibinfo{person}{Navid Rekabsaz}, \bibinfo{person}{Mihai
  Lupu}, \bibinfo{person}{Allan Hanbury}, {and} \bibinfo{person}{Guido
  Zuccon}.} \bibinfo{year}{2016}\natexlab{}.
\newblock \showarticletitle{Generalizing translation models in the
  probabilistic relevance framework}. In \bibinfo{booktitle}{{\em In Proc. of
  CIKM}}.
\newblock


\bibitem[\protect\citeauthoryear{Woodland, Johnson, Jourlin, and
  Jones}{Woodland et~al\mbox{.}}{2000}]%
        {woodland2000effects}
\bibfield{author}{\bibinfo{person}{Philip Woodland}, \bibinfo{person}{Sue
  Johnson}, \bibinfo{person}{Pierre Jourlin}, {and}
  \bibinfo{person}{Karen~Sp{\"a}rck Jones}.} \bibinfo{year}{2000}\natexlab{}.
\newblock \showarticletitle{Effects of out of vocab. words in spoken document
  retrieval}. In \bibinfo{booktitle}{{\em In Proc. of SIGIR}}.
\newblock


\bibitem[\protect\citeauthoryear{Xiong, Dai, Callan, Liu, and Power}{Xiong
  et~al\mbox{.}}{2017}]%
        {Xiong2017}
\bibfield{author}{\bibinfo{person}{Chenyan Xiong}, \bibinfo{person}{Zhuyun
  Dai}, \bibinfo{person}{Jamie Callan}, \bibinfo{person}{Zhiyuan Liu}, {and}
  \bibinfo{person}{Russell Power}.} \bibinfo{year}{2017}\natexlab{}.
\newblock \showarticletitle{{End-to-End Neural Ad-hoc Ranking with Kernel
  Pooling}}. In \bibinfo{booktitle}{{\em Proc. of SIGIR}}.
\newblock


\bibitem[\protect\citeauthoryear{Yang, Fang, and Lin}{Yang
  et~al\mbox{.}}{2017}]%
        {Yang2017}
\bibfield{author}{\bibinfo{person}{Peilin Yang}, \bibinfo{person}{Hui Fang},
  {and} \bibinfo{person}{Jimmy Lin}.} \bibinfo{year}{2017}\natexlab{}.
\newblock \showarticletitle{{Anserini: Enabling the use of Lucene for
  information retrieval research}}. In \bibinfo{booktitle}{{\em Proc. of
  SIGIR}}.
\newblock


\bibitem[\protect\citeauthoryear{Zamani, Dehghani, Croft, Learned-Miller, and
  Kamps}{Zamani et~al\mbox{.}}{2018}]%
        {zamani2018neural}
\bibfield{author}{\bibinfo{person}{Hamed Zamani}, \bibinfo{person}{Mostafa
  Dehghani}, \bibinfo{person}{W~Bruce Croft}, \bibinfo{person}{Erik
  Learned-Miller}, {and} \bibinfo{person}{Jaap Kamps}.}
  \bibinfo{year}{2018}\natexlab{}.
\newblock \showarticletitle{From Neural Re-Ranking to Neural Ranking: Learning
  a Sparse Representation for Inverted Indexing}. In \bibinfo{booktitle}{{\em
  In Proc. of CIKM}}.
\newblock


\end{thebibliography}

\end{document}